\pdfoutput=1

\documentclass[11pt]{article}

\usepackage{authblk}

\usepackage[preprint]{acl}
\usepackage{graphicx} 
\usepackage{svg}
\usepackage{subcaption}
\usepackage{array}
\usepackage{booktabs}
\usepackage{makecell, booktabs, caption}
\usepackage{footnote}  
\usepackage{amsmath} 
\usepackage{enumitem}
\usepackage[dvipsnames]{xcolor}
\usepackage{multirow}
\usepackage{times}
\usepackage{latexsym}
\usepackage{float}
\usepackage{natbib}

\usepackage{fvextra}
\usepackage{xspace}

\newcommand{\system}{\textsc{DCTR}\xspace}

\begin{document}

\title{Fine-Grained Table Retrieval Through the Lens of Complex Queries}

%
\author{\textbf{Wojciech Kosiuk}\textsuperscript{1},   
\textbf{Xingyu Ji}\textsuperscript{2}, \textbf{Yeounoh Chung}\textsuperscript{3},   \textbf{Fatma \"{O}zcan}\textsuperscript{3}, \textbf{Madelon Hulsebos}\textsuperscript{4}\thanks{Correspondence to: madelon@cwi.nl}\\
\textsuperscript{1}University of Amsterdam, \textsuperscript{2}UC Berkeley, \textsuperscript{3}Google, 
\textsuperscript{4}CWI
}

\maketitle

\begin{abstract}
\vspace{-0.15cm}
Enabling question answering over tables and databases in natural language has become a key capability in the democratization of insights from tabular data sources. These systems first require retrieval of data that is relevant to a given natural language query, for which several methods have been introduced. In this work we present and study a table retrieval mechanism devising fine-grained typed query decomposition and global connectivity-awareness (DCTR), to handle the challenges induced by open-domain question answering over relational databases in complex usage contexts. We evaluate the effectiveness of the two mechanisms through the lens of \textit{retrieval complexity} which we measure along the axes of query- and data complexity. Our analyses over industry-aligned benchmarks illustrate the robustness of DCTR for highly composite queries and densely connected databases.



\end{abstract}

\section{Introduction}
\vspace{-0.15cm}
Open-domain question answering and text-to-SQL over large heterogeneous table collections face a fundamental retrieval challenge: users rarely know which tables are relevant, and required join paths cannot be discerned from the query alone~\cite{chen2025tableretrievalsolvedproblem}. Ambiguous attribute names and inconsistent terminology create mismatches between natural-language queries and schema organization~\cite{lei-etal-2020-examining}. Multi-constraint queries allow for multiple plausible interpretations~\cite{gomm2025ambiguity}, making it difficult to identify correct tables before SQL generation. Retrieval errors propagate to downstream SQL generation, and are compounded when varying data model designs admit multiple valid SQL structures for the same query~\cite{fürst2024evaluatingdatamodelrobustness}, where even reliably evaluating query correctness remains challenging~\cite{kim2024flexexpertlevelfalselessexecution}. Moreover, while recent methods devise query decomposition for table retrieval, single-hop retrieval using full-query embeddings is a common default~\cite{chen2025enrichindexusingllmsenrich, zou2025gtr}, motivating a close examination of the robustness of single-hop retrieval for complex compositional queries.

Despite recent progress in table retrieval mechanisms, single-vector table embeddings cannot represent multi-constraint relevance signals~\cite{weller2025theoreticallimitationsembeddingbasedretrieval} and degrade when query semantics diverges from schema conventions~\cite{chen2025enrichindexusingllmsenrich, lei-etal-2020-examining}.
Industry workloads intensify these challenges: queries are verbose and compositional, databases span hundreds of tables with unaligned semantics, and relevant data is distributed across many join-connected tables. While existing studies surface patterns in retrieval performance across methods and datasets~\cite{ji2025targetbenchmarkingtableretrieval}, deeper insights are lacking regarding such complex retrieval characteristics.

Focusing on complex retrieval settings, we introduce Decomposition-based Connectivity Table Retrieval (DCTR) that 1) decomposes queries into semantic units of distinct types to enable fine-grained query-schema alignment using multi-vector embeddings, and 2) integrates global table connectivity-aware retrieval to surface relevant tables beyond semantic similarity. We evaluate \system{} on typical industry benchmarks in open-domain analytical tabular question answering which shows consistent gains in recall over common single-vector dense embedding retrieval. Finally, we formalize retrieval complexity along the query- and data dimensions, and analyze the influence of query signal-to-noise ratios, and schema heterogeneity and connectivity. Our findings surface the necessity of fine-grained query decomposition and global connectivity-awareness in complex retrieval settings involving highly compositional queries and large densely connected relational databases.

\section{Related Work}

\subsection{Text-to-SQL and Schema Linking}
Schema linking aligns natural-language queries with schema elements at varying granularities. RASL~\cite{eben2025raslretrievalaugmentedschema} introduces multi-stage retrieval for databases by decomposing schemas into semantic units. Nahid et al.~\cite{nahid2025rethinkingschemalinkingcontextaware} propose a context-aware bidirectional schema retrieval approach that links queries to tables and columns using LLM-guided selection. Recent work questions explicit linking: Maamari et al. ~\cite{maamari2024deathschemalinkingtexttosql} argue that high-capacity LLMs identify relevant schema despite distracting context, while RESDSQL~\cite{li2023resdsqldecouplingschemalinking} frames linking as ranking-enhanced encoding decoupled from SQL generation. These works generally assume a known data context and limited retrieval scope, providing a narrow view of the open-domain setting in complex data context.

\subsection{Retrieval in Open-Domain Text-to-SQL}
Open-domain QA faces ambiguous intent and heterogeneous corpora. MURRE~\cite{zhang2024murre} performs iterative multi-hop retrieval with query rewriting which is complemented by data augmentation in Abacus-SQL~\cite{xu2025abacus}, while LinkAlign~\cite{wang2025linkalign} scales schema linking to large databases via multi-round query rewriting and response filtering. EnrichIndex~\cite{chen2025enrichindexusingllmsenrich} and Pneuma~\cite{balaka2025pneuma} show the value of enriching retrieval indices with LLM-generated metadata, while GenEdit~\cite{maamari2025genedit} integrates domain knowledge in SQL generation through specialized operators and user feedback. Complementary techniques include graph-based hierarchical retrieval~\cite{zou2025gtr} and weak supervision~\cite{liang2025improvingtableretrievalquestion}. To address the large scale of databases, R2D2~\cite{DBLP:conf/aidm/BodensohnB24} retrieves table segments instead of full tables. JAR~\cite{chen2025tableretrievalsolvedproblem} introduces join-aware reranking to balance table-query relevance with table-table compatibility, which was adapted for iterative retrieval by \citet{boutaleb2025exploring}. We build on these principles, but introduce \textit{global} connectivity-aware retrieval to surface relevant tables beyond semantic similarity, and typed query decomposition for fine-grained schema alignment. We analyze the effectiveness through the lens of complex queries and databases providing a novel view of performance in complex settings.



\section{Problem Setting}






\paragraph{Open-Domain Table Retrieval}\label{par:retrieval-task-cr}
Let $T = \{T_1, \dots, T_n\}$ denote the set of tables in a relational database, where each table $T_i$ consists of a set of columns $C(T_i) = \{c_1, \dots, c_m\}$, possibly connected through foreign-key relationships. In an open-domain setting, a natural-language query $q$ expresses an information need from which the data scope needs to be inferred for retrieval of relevant tables from a large topic-diverse set of tables. The ground-truth set of relevant tables for $q$ is denoted as $T^*(q)$.  
The table retrieval task aims to return a ranked list $R_k(q)$ of top-$k$ candidate tables such that capped recall (CR)~\cite{thakur2beir} is maximized: $\mathrm{CR@k}(q) = \frac{|R_k(q) \cap T^*(q)|}{\min(k, |T^*(q)|)}$.
\vspace{-0.1cm}

\paragraph{Retrieval Complexity}\label{par:retrieval-complexity}

The composition of queries and data that is typically faced in real-world retrieval tasks for end-to-end (analytical) question answering stretches the complexity levels of what existing evaluations surface. To serve inspection from this angle, we characterize retrieval complexity along two dimensions: \textit{query complexity} and \textit{data complexity}. We term \underline{query complexity} as \textit{the complexity induced by how the query is formulated in terms of semantic density and functional composition}~\cite{gomm2025ambiguity}. We term \underline{data complexity} as \textit{the complexity resulting from how the data is structured within a database due to database normalization and schema size}~\cite{chen2025tableretrievalsolvedproblem}. These complexity dimensions guide our analysis of fine-grained retrieval for complex question answering tasks.
\vspace{-0.1cm}



\section{Fine-Grained Table Retrieval}

We describe the Decomposition and Connectivity-aware Table Retrieval method (\system{}) that is tailored for complex retrieval tasks over two dimensions: 1) functional query decomposition for fine-grained schema alignment, and 2) global connectivity aware retrieval for recovering relevant tables beyond the semantic similarity scope.

Before queries are executed, an index of the data is built first. To support scalable retrieval over large schemas with thousands of tables and a proportional number of columns, we construct two dense indices: one encoding table names and one encoding column names. Retrieval is performed by comparing embeddings of query elements against embeddings of these different schema elements.

\subsection{\system{} Retrieval Pipeline}

\begin{figure}[th]
    \centering
    \includegraphics[width=\linewidth]{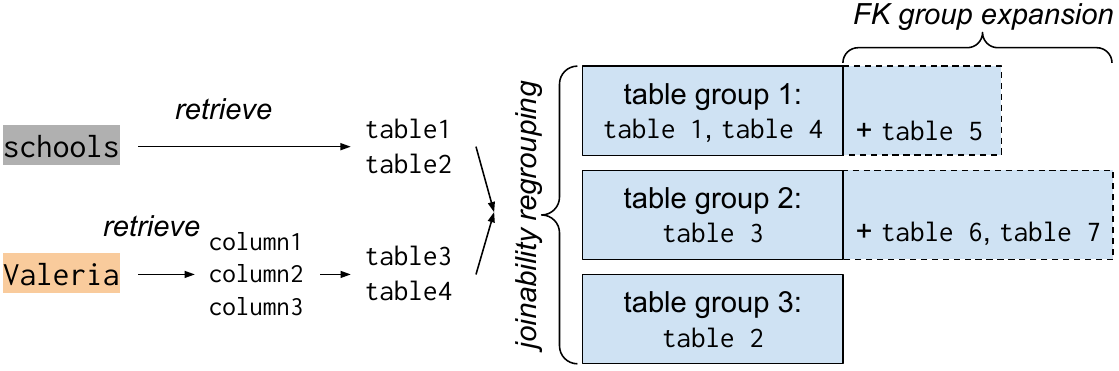}
    \caption{Overview of DCTR with component-wise retrieval, group formation, and FK-group expansion.}
    \label{fig:methodology:onlineretrieval}
    \vspace{-0.2cm}
\end{figure}

Figure~\ref{fig:methodology:onlineretrieval} shows the retrieval pipeline at inference from component extraction and component-wise retrieval, to join-graph construction and FK-based expansion to account for global connectivity.

\paragraph{Typed Query Decomposition in Semantic Units}
\label{subsec:query_decomposition}
In an open-domain setting, natural language queries reflect constraints over different entities, attributes, and operations~\cite{gomm2025ambiguity}. To align these with concrete schema elements and operators, we first perform a \emph{typed query decomposition} step that segments an incoming query $q$ into atomic \emph{semantic units}, forming the basis for both retrieval and downstream task execution (e.g. SQL generation). The decomposition is done using a language model with a fixed few-shot instruction prompt and decoding configuration to ensure consistent behavior across datasets (Appendix~\ref{app:query-prompt}). \system{} classifies each \textit{semantic unit} as one of three component types:
\begin{itemize}[topsep=3pt, itemsep=1pt, parsep=3pt]
    \item \colorbox{Gray!80}{Schema components}, denoting candidate tables or columns names and are matched to database schemas during retrieval.
    \item \colorbox{Dandelion!80}{Value components}, corresponding to entities or literals that act as query filters.
    \item \colorbox{Lavender!80}{Aggregator components}, which specify aggregation or comparison operations (e.g., \texttt{average}, \texttt{min}, \texttt{max}).
\end{itemize}

For example, the query ``\textit{What was the average sale of Luka Dončić jerseys in 2025?}'' is decomposed into  schema components \colorbox{Gray!80}{\texttt{jersey}} and \colorbox{Gray!80}{\texttt{sale}}, value components \colorbox{Dandelion!80}{\texttt{Luka Dončić}} and \colorbox{Dandelion!80}{\texttt{2025}}, and aggregator component \colorbox{Lavender!80}{\texttt{average}}. In the retrieval phase, only schema and value components are used while the aggregator components benefits the downstream tasks, e.g. text-to-SQL.

\paragraph{First Pass Retrieval}
In the first stage of retrieval, each query component is processed independently for each database. For each component, we perform dense similarity search against both the table and column indices, retrieving the
candidates per component. This initial retrieval produces unranked lists of tables and columns. The results from all components and all databases are then merged into a single candidate set. Columns retrieved from the column index are mapped back to their parent tables, ensuring that all relevant tables are included in the set that is forwarded to the graph-based grouping stage.

\paragraph{Schema Graph and Group Construction}
\label{subsubsec:graph-groups}
As discussed in \citep{chen2025tableretrievalsolvedproblem}, joinable tables are relevant to consider in table retrieval but might not necessarily surface in a first-pass retrieval stage which is conditioned on semantic similarity with the input query components hence require implicit multi-hop reasoning over join paths. Given the set of candidate tables retrieved in the first pass, we represent each database schema as an undirected graph where nodes correspond to tables and edges represent foreign-key join relationships.

We restrict the global schema graph to nodes corresponding to the retrieved candidate tables and consider only valid foreign-key (FK) join relationships between them. This subgraph captures the joinability structure among candidate tables relevant to the query
while excluding unrelated parts of the schema. Each connected component defines a \emph{table group}. By global FK-expansion, we expand each table group by including all tables that are connected through FKs with any table in the group, but which are not surfaced through semantic similarity, recovering multi-hop join context.



\paragraph{Group Scoring}
\label{subsubsec:group-scoring}

Each table group is scored based on its coverage of the extracted query components.  
For each component, \system{} first computes the cosine similarity between its embedding and the embeddings of all tables in the group. 

We introduce a parameter \textit{vote\_k} to cap the number of tables included in the scoring of a table group. The resulting \emph{group coverage score} is obtained by summing the similarity scores of the top \textit{vote\_k} tables per component across all components: $\text{group\_cov\_score(T, C)}=\sum_{j=1}^{|C|} \text{top}_{\text{vote}_k}
(\{ \text{sim}(t_{1}, C_{j}), ..., \text{sim}(t_{n}, C_{j}) \})$.


Higher group coverage score indicates that the tables in the group collectively cover more of the query's components, and therefore are more likely to contain the information needed to answer the query accurately. In the current formulation, all components contribute equally to the group score.

\paragraph{Final Retrieval}
\label{subsec:finalretrieval}

The final retrieval stage selects the most relevant tables by retaining the top $n\_{\text{groups}}$ scoring table groups per database. Within each selected group, the $\text{vote}\_k$ tables with the highest individual coverage scores are selected as the final candidate set. These two hyperparameters control the breadth and granularity of the retrieval, allowing the pipeline to balance recall and precision depending on query complexity and schema size.\\

\section{Experimental Setup}

\subsection{Method and Baselines Implementation}

\paragraph{Retrieval Implementation} We implement the retrieval backend using Qdrant with two collections. Schema representations are indexed separately: table identifiers are stored in \texttt{table\_embeddings}, while column identifiers and their semantic-type variants are stored in \texttt{column\_embeddings}. For query decomposition we use \texttt{gpt-4.1-mini} and \texttt{gemini-2.5-pro} for text-to-SQL evaluation.

\paragraph{Hyperparameters}\label{subsec:hyperparameters}
\system{} introduces three key hyperparameters we evaluate across experiments:

\begin{itemize}[topsep=5pt, leftmargin=15pt, itemsep=3pt, parsep=5pt]
    \item \text{vote\_$k$}: Number of in-group candidates selected per component for voting during group scoring.
    \item \text{$n$\_groups}: Number of table subgroups retained per database for downstream ranking.
    \item \text{expand\_groups}: Whether to expand high-scoring subgroups via join paths in the schema graph before scoring.
\end{itemize}


\paragraph{Dense Retrieval Baseline}\label{subsec:dense-baseline}

We compare our method against the common dense retrieval baseline \cite{karpukhin2020densepassageretrievalopendomain}. Given a natural-language query $q$, the baseline embeds full query as a single vector and tables are represented by an embedding of their name. Retrieval is performed based on the highest cosine similarity across table embeddings within each database and the given query embedding.

\paragraph{Embedding Models}

All retrieval components use dense text embeddings for query units and schema elements. We evaluate the following three models covering different capacity–efficiency trade-offs. \textbf{Stella-large}~\cite{zhang2025jasperstelladistillationsota}: a high-capacity model (400M parameters, 1024-dim) designed for fine-grained semantic representation. \textbf{BGE-small}~\cite{xiao2024cpackpackedresourcesgeneral}: a lightweight general-purpose model (33M parameters, 384-dim) serving as an efficiency-oriented baseline. \textbf{E5-small}~\cite{wang2024textembeddingsweaklysupervisedcontrastive}: A lightweight model (33M parameters, 384-dim) trained with weakly supervised contrastive objectives for scalable retrieval.

\subsection{Datasets and Metrics}

\paragraph{Datasets}
We evaluate on three text-to-SQL benchmarks accessed via 
TARGET~\cite{ji2025targetbenchmarkingtableretrieval}, 
selected to mirror the friction points of 
industry retrieval: large schemas, 
ambiguous column naming, compositional queries, 
and multi-level nested joins. Together they span enterprise settings, 
stretching both query and data complexity (Sec.~\ref{par:complexity}).
\textbf{BEAVER}~\cite{chen2025beaverenterprisebenchmarktexttosql}: 
an enterprise benchmark from real data warehouses, 
with large schemas (77.2 tables per database on average) 
and verbose queries (31.9 words on average) involving 
multi-table joins and aggregations.
\textbf{FIBEN}~\cite{10.14778/3407790.3407858}: a finance-domain 
benchmark with 300 natural-language questions over a single 
dense schema of 152 tables, requiring reasoning over 
domain-specific jargon and complex nested queries.
\textbf{BIRD}~\cite{li2023llmservedatabaseinterface}: a 
cross-domain benchmark spanning 11 databases with smaller 
but well-connected schemas (6.8 tables, 1.4 FKs per table).
We report results on the test sets of BEAVER and FIBEN, 
and the validation set of BIRD. Full dataset characteristics 
are presented in Table~\ref{tab:dataset-characteristics} 
(Appendix~\ref{appendix:distribution}).

\paragraph{Retrieval Metrics} We evaluate the proposed table retrieval pipeline using Capped Recall RC@\(k\) for \(k \in \{5,10,25\}\) across all embedding models and benchmarks (Sec.~\ref{par:retrieval-task-cr}). First-stage candidate breadth is fixed to 30, and hyperparameters are tuned to maximize capped Recall@\(k\). Results are reported over multiple runs to account for stochasticity from LLM query decomposition. The dense baseline is deterministic and shows no variance.

\paragraph{Complexity Measures} \label{par:complexity} Following our notion of retrieval complexity as defined in Sec. \ref{par:retrieval-complexity}, we investigate retrieval performance under varying degrees of \textit{query complexity} and \textit{data complexity}, as typical industry workloads stretch these attributes. For each query, we capture attributes outlined below and analyze capped recall@25, aggregated across datasets and embedding models.

\clearpage

For \emph{query complexity}, we measure 1) the length of the query in number of tokens, and 2) the number of extracted functional query components. For \emph{data complexity}, we measure the total number of tables connected through FKs with the gold tables involved in a query reflecting the degree of schema normalization and connectivity. Distributional statistics of the datasets along these dimensions are presented in Appendix~\ref{appendix:distribution}.


\section{Results}


\subsection{Overall Retrieval Performance} Tables \ref{tab:recall_k5}-\ref{tab:recall_k25} report capped Recall@$k$ for $k \in \{5,10,25\}$ across datasets, embedding models, and retrieval settings. For small retrieval scope ($k=5$), small settings for $\text{vote}\_k$ and $n\_\text{groups}$ perform best. As $k$ increases, moderate $\text{vote}\_k$ and $n\_\text{groups}$ improve recall over the dense baseline, with the largest gains on BEAVER and for smaller embedding models. Notably, small embedding models benefit most from \system{}, narrowing the gap with a high-capacity model in the baseline.BEAVER reflects industry settings with larger schemas, whereas BIRD consists of smaller, isolated databases. Structured, connectivity-aware retrieval is particularly beneficial in enterprise-like settings for retrieving tables from larger schemas. Good priors for \system{} settings are proportional to the retrieval scope: for small $k$ (e.g. 5) increasing $\text{vote}_k$ or $n_{\text{groups}}$ is ineffective, while for larger $k$, i.e. 10 or 25, setting $\text{vote}_k$ and $n_{\text{groups}}$ to 2 and 5, respectively, yields consistent gain in recall (Appendix~\ref{appendix:ablation}).

Group expansion shows dataset-dependent behavior. It improves recall on BIRD (+3–5\%), particularly at larger $k$, but substantially degrades performance on FIBEN and BEAVER (up to 30\%). This aligns with structural differences across benchmarks (Appendix~\ref{appendix:distribution} Table~\ref{tab:dataset-characteristics}). That is, BIRD databases are small (6.8 tables on average) yet well-connected (1.4 FKs per table), making expansion effective for recovering non-semantically similar but relevant tables. In contrast, FIBEN and BEAVER contain substantially larger schemas (respectively 152 and 77.5 tables per database), yielding significant more candidate tables that might get truncated by the $k$-capped evaluation. These dataset characteristics suggest that a dynamic value for $k$ would be viable in practice to fully utilize global FK-expansion.


\begin{table}[th]
\centering
\caption{Capped Recall@5 (mean $\pm$ std) across datasets and models. Bold values indicate the best result per dataset and embedding model.}\label{tab:recall_k5}
\renewcommand{\arraystretch}{1.2}
\resizebox{\columnwidth}{!}{%
    \begin{tabular}{llccc}
    \toprule
    \textbf{Setting} & \textbf{Model} & \textbf{BEAVER} & \textbf{FIBEN} & \textbf{BIRD} \\
    \midrule
    \multirow{3}{2cm}{Baseline}
     & stella    & 0.181 & 0.297 & \textbf{0.901} \\
     & bge-small    & 0.154 & 0.198& 0.809 \\
     & e5-small     & 0.108 & 0.062 & 0.737 \\
    \midrule
    \multirow{3}{2cm}{$\text{vote}_k\!=\!1$, $n_{\text{groups}}\!=\!1$}
     & stella    & \textbf{0.231}{\tiny$\pm$.004} & \textbf{0.317}{\tiny$\pm$.001} & 0.834{\tiny$\pm$.001} \\
     & bge-small    & \textbf{0.223}{\tiny$\pm$.005} & \textbf{0.294}{\tiny$\pm$.003} & \textbf{0.826}{\tiny$\pm$.002} \\
     & e5-small     & \textbf{0.199}{\tiny$\pm$.009} & \textbf{0.257}{\tiny$\pm$.004} & \textbf{0.817}{\tiny$\pm$.007} \\
    \bottomrule
    \end{tabular}
}
\vspace{0.2cm}
\end{table}

\begin{table}[th]
\centering
\caption{Capped Recall@10 (mean $\pm$ std) across datasets and models. Bold values indicate the best result per dataset and embedding model.}\label{tab:recall_k10}
\renewcommand{\arraystretch}{1.2}
\resizebox{\columnwidth}{!}{%
    \begin{tabular}{llccc}
    \toprule
    \textbf{Setting} & \textbf{Model} & \textbf{BEAVER} & \textbf{FIBEN} & \textbf{BIRD} \\
    \midrule
    \multirow{3}{2cm}{Baseline}
     & stella    & 0.277 & \textbf{0.413} & \textbf{0.961} \\
     & bge-small    & 0.231 & 0.262 & 0.909 \\
     & e5-small     & 0.175 & 0.140 & 0.850 \\
    \midrule
    \multirow{3}{2cm}{$\text{vote}_k\!=\!2$, $n_{\text{groups}}\!=\!2$}
     & stella    & \textbf{0.313}{\tiny$\pm$.007} & 0.396{\tiny$\pm$.006} & 0.899{\tiny$\pm$.004} \\
     & bge-small    & \textbf{0.326}{\tiny$\pm$.002} & \textbf{0.331}{\tiny$\pm$.006} & 0.893{\tiny$\pm$.003} \\
     & e5-small     & \textbf{0.273}{\tiny$\pm$.007} & \textbf{0.346}{\tiny$\pm$.004} & 0.882{\tiny$\pm$.006} \\
    \midrule
    \multirow{3}{2cm}{$\text{vote}_k\!=\!2$, $n_{\text{groups}}\!=\!2$, $\text{expand\_groups}$}
     & stella    & 0.073{\tiny$\pm$.014} & 0.202{\tiny$\pm$.002} & 0.936{\tiny$\pm$.003} \\
     & bge-small    & 0.095{\tiny$\pm$.001} & 0.091{\tiny$\pm$.014} & \textbf{0.942}{\tiny$\pm$.001} \\
     & e5-small     & 0.109{\tiny$\pm$.001} & 0.078{\tiny$\pm$.008} & \textbf{0.904}{\tiny$\pm$.005} \\
    \bottomrule
    \end{tabular}
}
\end{table}


\begin{table}[th]
\centering
\caption{Capped Recall@25 (mean $\pm$ std) across datasets and models. Bold values indicate the best result per dataset and embedding model.}\label{tab:recall_k25}
\vspace{-0.2cm}
\renewcommand{\arraystretch}{1.2}
\resizebox{\columnwidth}{!}{%
    \begin{tabular}{llccc}
        \toprule
        \textbf{Setting} & \textbf{Model} & \textbf{BEAVER} & \textbf{FIBEN} & \textbf{BIRD} \\
        \midrule
        \multirow{3}{2.1cm}{Baseline}
         & stella    & 0.413 & \textbf{0.584} & 0.986 \\
         & bge-small    & 0.350 & \textbf{0.461} & 0.976 \\
         & e5-small     & 0.281 & 0.283          & 0.949 \\
        \midrule
        \multirow{3}{2.1cm}{$\text{vote}_k\!=\!5$, $n_{\text{groups}}\!=\!5$}
         & stella    & \textbf{0.435}{\tiny$\pm$.002} & 0.513{\tiny$\pm$.007} & 0.971{\tiny$\pm$.002} \\
         & bge-small    & \textbf{0.429}{\tiny$\pm$.003} & 0.448{\tiny$\pm$.001} & 0.972{\tiny$\pm$.002} \\
         & e5-small     & \textbf{0.408}{\tiny$\pm$.008} & \textbf{0.516}{\tiny$\pm$.003} & 0.965{\tiny$\pm$.001} \\
        \midrule
        \multirow{3}{2.1cm}{$\text{vote}_k\!=\!5$, $n_{\text{groups}}\!=\!5$, $\text{expand\_groups}$}
         & stella    & 0.228{\tiny$\pm$.001} & 0.259{\tiny$\pm$.001} & \textbf{0.991}{\tiny$\pm$.000} \\
         & bge-small    & 0.232{\tiny$\pm$.002} & 0.127{\tiny$\pm$.014} & \textbf{0.988}{\tiny$\pm$.002} \\
         & e5-small     & 0.228{\tiny$\pm$.001} & 0.011{\tiny$\pm$.003} & \textbf{0.975}{\tiny$\pm$.001} \\
        \bottomrule
    \end{tabular}
    }
\end{table}

\subsection{Retrieval Complexity Analysis}
\label{subsec:perquery_experiments}

\begin{figure}[th]
\centering
    \includesvg[width=\linewidth]{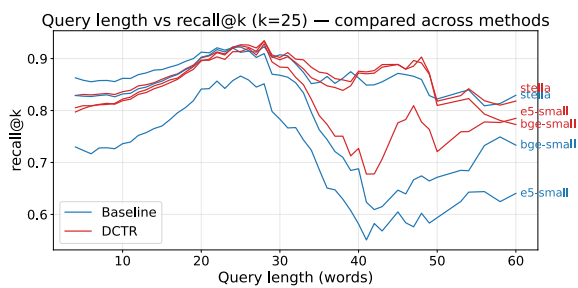}
     \vspace{-0.5cm}
  \caption{Query length vs recall@25, aggregated across datasets. Comparison through three embedding models, shows that \system{} outperforms the baseline, especially for longer queries ($\geq$40 tokens). 
  }
  \label{fig:qlen}
  \vspace{-0.5cm}
\end{figure}

\paragraph{Query complexity.}
Figure~\ref{fig:qlen} shows the retrieval performance against \textit{query length} in number of tokens. We find that \system{} consistently outperforms the dense baseline for longer queries, with the gap widening beyond 40 tokens. For smaller embedding models, in particular, the baseline is more sensitive to noise in the query than \system{} illustrating the effectiveness of fine-grained query decomposition for complex queries.

\begin{figure}[th]
  \centering
    \includesvg[width=\linewidth]{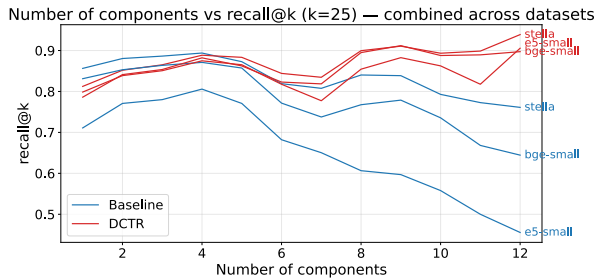}
    \vspace{-0.7cm}
  \caption{Number of components vs Recall@25 across datasets. Comparison through three embedding models shows that \system{} is robust for compositional queries while a single-vector baseline does not resolve compositional queries regardless of model capacity.}
  \label{fig:components}
\end{figure}

When considering retrieval performance over the number of extracted components per query (Fig.~\ref{fig:components}), we observe that both methods perform comparable for lightly compositional queries but the baseline's recall degrades steadily as the number of components increases while \system{} remains stable. The decay of the baseline, for all embedding models, highlights that high-capacity models cannot resolve highly composite queries, while they are more robust to irrelevant noise (Fig.~\ref{fig:qlen}). This confirms the need for fine-grained, component-level, retrieval for highly compositional queries.

\paragraph{Data complexity.}

\begin{figure}[th]
  \centering
    \includesvg[width=\linewidth]{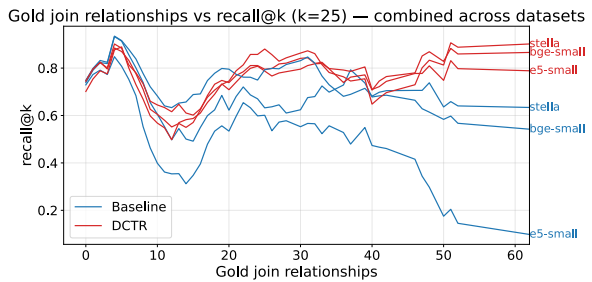}
     \vspace{-0.7cm}
  \caption{Number of tables connected to gold tables vs recall@25, combined across datasets, for different embedding models. \system{} improves more on queries where gold tables are densely connected.}
  \label{fig:num_joins}
  \vspace{-0.1cm}
\end{figure}

We assess data complexity via the connectivity of gold tables, measured by their number of join relationships within the database. Figure~\ref{fig:num_joins} illustrates that \system{} achieves stronger performance on queries involving densely connected tables, indicating that connectivity-aware grouping improves retrieval of joinable table sets.

\subsection{Downstream Text-to-SQL Performance}
\label{subsec:exp_ttsql}
We evaluate the effectiveness of \system{} on downstream text-to-SQL to assess practical utility of retrieval versus using long-context in complex retrieval tasks. We pass the full schema as context to Gemini for the baseline~\cite{Chung_2025}, or the retrieved tables using \system{} with \textit{stella} embeddings and $k \in \{5, 10, 25\}$. Performance is measured by execution accuracy (EX). Results are reported in Table~\ref{tab:sql}. \system{} improves execution accuracy by +3\% on BEAVER and +5\% on FIBEN, while matching the baseline on BIRD. The performance varies over $k$ per dataset and correlates with retrieval difficulty: BIRD achieves peak accuracy at small $k$ where recall is already high, FIBEN requires moderate $k$, and BEAVER, where retrieval is most challenging, benefits from larger $k$.

\begin{table}[h!]
\centering
\footnotesize
\caption{Downstream SQL execution accuracy of \system{} against the long-context baseline.
}
\vspace{-0.3cm}
\label{tab:sql}
\makebox[\columnwidth][c]{%
\begin{tabular}{>{\raggedright\arraybackslash}p{1.4cm} l c c c}
\toprule
\textbf{Setting} & & \textbf{BEAVER} & \textbf{FIBEN} & \textbf{BIRD} \\
\midrule
Baseline
&  & 0.2730 & 0.1370 & 0.6630\\
\midrule
\system{}
& k=5     & 0.2150 & 0.0970 & \textbf{0.6656} \\
& k=10  & 0.2400 & \textbf{0.1870} & 0.6630 \\
& k=25   & \textbf{0.2980} & 0.1270 & 0.6623\\
\bottomrule
\end{tabular}}
\vspace{-0.3cm}
\end{table}

\vspace{-0.1cm}
\section{Conclusion}
\vspace{-0.2cm}






The emergence of natural language interfaces to large-scale databases amplifies the importance of retrieval mechanisms that are robust to complex query- and data characteristics. In this context, we introduce Decomposition-based Connectivity-aware Table Retrieval (\system{}) which, we show, has significant gains compared to common single-vector dense retrieval baselines in industry-representative benchmarks. Analysis of retrieval complexity along the axes of query- and data complexity surfaces that typed query decomposition is key in addressing highly compositional queries that require fine-grained multi-table schema alignment, while global connectivity-aware retrieval uncovers relevant tables beyond semantically similar tables in densely connected relational databases. 

\bibliography{main.bib}


\appendix
\section*{Appendix}





\section{Query Decomposition Prompt}
\label{app:query-prompt}

We use a fixed instruction prompt to extract schema-related query components.
The prompt is shown below:

\small
\begin{Verbatim}[breaklines=true]
---Goal---
Given a natural language query and a list of component types, identify all schema
element components from the text.

---Output---
For each identified component, extract the following information:
- component_name: Name of the component, use exact phrase from the input query.
  Keep each as short and concise as possible.
- component_description: Short description of the component's attributes and any
  potential context that might be helpful for understanding the role of the component
  within the query.

---Schema Element Component Description---
The specific data schema element (e.g. table or column names) directly referenced
in the query. Think about which parts of the query are most likely to have counterparts
in the available data, and explain why. Be as concise as possible, and shave any
constraints or filtering phrases surrounding the schema element component.

Note: Exclude phrases that are more likely to refer to specific cell values rather than
schema elements. Only focus on elements most likely to have matching schema entries.

---Examples---
{few_shot_examples}

---Real Data---
Input Query:
{query}

Output:
\end{Verbatim}

\normalsize

\section{Ablation Study}
\label{appendix:ablation}

We study the effect of key hyperparameters (\textit{n\_groups}, \textit{vote\_k}, and \textit{expand\_groups}) on retrieval performance. Since datasets differ in database count, schema size, and inter-table connectivity, hyperparameter selection is essential for balancing coverage and precision. We focus on \(k=25\), where the effects of grouping and voting are most pronounced and easier to observe.


\paragraph{Results.}
Table~\ref{tab:hyperparameter_ablation} summarizes the ablation results. Across datasets, increasing \textit{n\_groups} and \textit{vote\_k} consistently improves recall up to a saturation point, beyond which gains diminish or disappear. For \(k=25\), optimal performance is typically achieved with \textit{n\_groups} and \textit{vote\_k} in the range of 5–6, while larger values yield no further improvement.

On BEAVER and FIBEN, enabling group expansion substantially degrades performance across all configurations. In contrast, BIRD benefits from group expansion in several settings, often reaching near-perfect recall. These trends are consistent across embedding models, indicating that hyperparameter behavior is largely independent of the embedding choice and instead driven by dataset characteristics.


\begin{table*}[th]
\centering
\footnotesize
\setlength{\tabcolsep}{3pt}
\makebox[\columnwidth][c]{%
\begin{tabular}{>{\raggedright\arraybackslash}p{1cm} c|ccc|ccc}
\toprule
\textbf{Dataset} & \textbf{$(n_{\text{groups}},\, vote_k)$} &
\multicolumn{6}{c}{\textbf{Embedding Model}} \\
\cmidrule(lr){3-8}
& &
\rotatebox{45}{\textbf{stella}} &
\rotatebox{45}{\textbf{bge-small}} &
\rotatebox{45}{\textbf{e5-small}} &
\rotatebox{45}{\textbf{stella}} &
\rotatebox{45}{\textbf{bge-small}} &
\rotatebox{45}{\textbf{e5-small}} \\
& &
\multicolumn{3}{c|}{\textbf{expand\_groups=False}} &
\multicolumn{3}{c}{\textbf{expand\_groups=True}} \\
\midrule
BEAVER & (3,3) & 0.3901 & 0.3869 & 0.3782 & 0.2231 & 0.2338 & 0.2155 \\
       & (4,4) & 0.4106 & 0.4073 & 0.3933 & 0.2306 & 0.2392 & 0.2392 \\
       & (5,5) & 0.4332 & 0.4321 & 0.4106 & 0.2284 & 0.2306 & 0.2284 \\
       & (6,6) & 0.4504 & 0.4332 & 0.4138 & 0.2285 & 0.2349 & 0.2371 \\
       & (7,7) & 0.4569 & 0.4353 & 0.3966 & 0.2381 & 0.2403 & 0.2371 \\
\midrule
FIBEN & (3,3) & 0.4526 & 0.3890 & 0.4091 & 0.2324 & 0.1149 & 0.1227 \\
                & (4,4) & 0.4769 & 0.4051 & 0.4508 & 0.2524 & 0.1271 & 0.1044 \\
                & (5,5) & 0.5152 & 0.4491 & 0.5178 & 0.2594 & 0.1366 & 0.1131 \\
                & (6,6) & 0.5466 & 0.4700 & 0.5535 & 0.2672 & 0.1271 & 0.1262 \\
                & (7,7) & 0.5648 & 0.4795 & 0.5614 & 0.2715 & 0.1393 & 0.1184 \\
\midrule
BIRD\\(200 queries) & (3,3) & 0.9214 & 0.9337 & 0.9509 & 1.0000 & 1.0000 & 0.9951 \\
               & (4,4) & 0.9582 & 0.9754 & 0.9730 & 1.0000 & 1.0000 & 0.9975 \\
               & (5,5) & 0.9754 & 0.9705 & 0.9902 & 1.0000 & 1.0000 & 0.9951 \\
               & (6,6) & 0.9754 & 0.9803 & 0.9926 & 1.0000 & 1.0000 & 0.9877 \\
               & (7,7) & 0.9754 & 0.9779 & 0.9926 & 1.0000 & 0.9926 & 0.9975 \\
\bottomrule
\end{tabular}}
\caption{Hyperparameter ablation results. Capped Recall@25 for different $(n_{\text{groups}},\, vote_k)$ settings.}
\label{tab:hyperparameter_ablation}
\end{table*}

\paragraph{Ablation results for k=5,10}
\label{appendix:ablation510}
For $k \in {5,10}$, we evaluate larger hyperparameter configurations $(n_{\text{groups}}, \text{vote}_k)$, with settings 1–3 considered for $k=5$ and 1–4 for $k=10$, each evaluated both with and without \textit{expand\_groups}. 

Results in Table~\ref{tab:hyperparameter_ablation_stella} show the effect of hyperparameter choices for $k=5$ and $k=10$, using only \texttt{stella-large}, as no significant differences were observed across models.

The optimal $(n_{\text{groups}},, vote_k)$ configuration varies by dataset. For BEAVER, the best performance is achieved with $(1,1)$ at $k=5$ and $(2,2)$ at $k=10$. In FIBEN, $(1,1)$ yields the highest recall for both $k=5$ and $k=10$. For BIRD, $(3,3)$ provides the best results at $k=5$. Note that BIRD is evaluated on a 200-query subsample, so these optimal values differ when considering the full dataset.

\begin{table}[th]
\centering
\footnotesize
\setlength{\tabcolsep}{4pt}
\makebox[\columnwidth][c]{%
\begin{tabular}{>{\raggedright\arraybackslash}p{1.2cm} c|cc|cc}
\toprule
\textbf{Dataset} & \textbf{$(n_{\text{groups}},\, vote_k)$} &
\multicolumn{2}{c|}{\textbf{Recall@5}} &
\multicolumn{2}{c}{\textbf{Recall@10}} \\
\cmidrule(lr){3-4} \cmidrule(lr){5-6}
& & False & True & False & True \\
\midrule
BEAVER & (1,1) & 0.2333 & 0.0317 & 0.2575 & 0.0830 \\
       & (2,2) & 0.1993 & 0.0294 & 0.3308 & 0.0873 \\
       & (3,3) & 0.1880 & 0.0294 & 0.3114 & 0.0862 \\
       & (4,4) &      &      & 0.2931 & 0.0916 \\
\midrule
FIBEN & (1,1) & 0.4626 & 0.1737 & 0.4743 & 0.1766 \\
       & (2,2) & 0.3763 & 0.1694 & 0.4034 & 0.2066 \\
       & (3,3) & 0.3477 & 0.1604 & 0.4368 & 0.1988 \\
       & (4,4) &      &      & 0.4359 & 0.1918 \\
\midrule
BIRD\\(200 queries) & (1,1) & 0.8354 & 0.7641 & 0.8452 & 0.9165 \\
       & (2,2) & 0.8697 & 0.8010 & 0.9042 & 0.9754 \\
       & (3,3) & 0.8747 & 0.7740 & 0.9140 & 0.9754 \\
       & (4,4) &      &      & 0.9435 & 0.9779 \\
\bottomrule
\end{tabular}}
\caption{Hyperparameter ablation results for \texttt{stella-large} embedding model. Capped Recall@5 and @10 are reported for different $(n_{\text{groups}},\, vote_k)$ settings, with \textit{expand\_groups} enabled (True) or disabled (False).}
\label{tab:hyperparameter_ablation_stella}
\end{table}

\section{Results for Detailed Per-Query Analysis for k=5,10}
\label{appendix:per_query_results}

\begin{figure}[th]
\centering

  \begin{subfigure}{\linewidth}
    \centering
    \includesvg[width=0.7\linewidth]{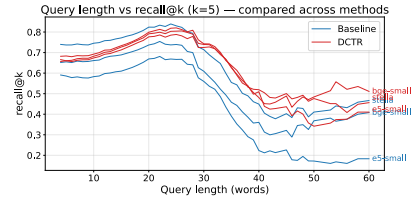}
    \caption{CappedRecall@5}
  \end{subfigure}

  \vspace{0.5em}

  \begin{subfigure}{\linewidth}
    \centering
    \includesvg[width=0.7\linewidth]{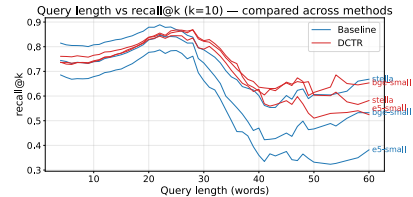}
    \caption{CappedRecall@10}
  \end{subfigure}

  \vspace{0.5em}

  \begin{subfigure}{\linewidth}
    \centering
    \includesvg[width=0.7\linewidth]{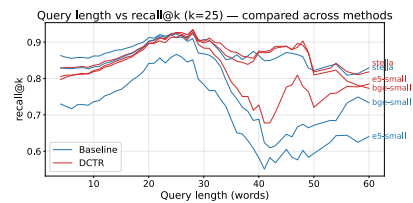}
    \caption{CappedRecall@25}
  \end{subfigure}
  \caption{Query length vs recall@k, combined across datasets. Comparison of three embedding models for the proposed method versus the baseline. Across all values of $k$, the proposed method consistently outperforms the baseline, especially for longer queries ($\geq$40 tokens).}
  \label{fig:qlen-combined}
\end{figure}

\begin{figure}[th]
  \centering

  \begin{subfigure}{\linewidth}
    \centering
    \includesvg[width=0.7\linewidth]{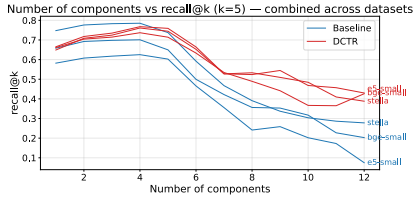}
    \caption{CappedRecall@5}
  \end{subfigure}

  \vspace{0.5em}

  \begin{subfigure}{\linewidth}
    \centering
    \includesvg[width=0.7\linewidth]{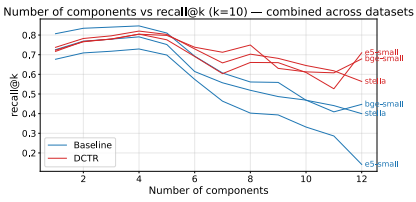}
    \caption{CappedRecall@10}
  \end{subfigure}

  \vspace{0.5em}

  \begin{subfigure}{\linewidth}
    \centering
    \includesvg[width=0.7\linewidth]{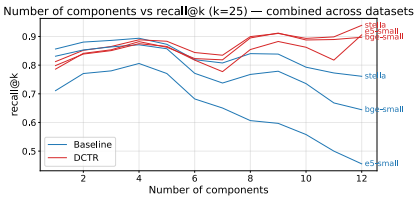}
    \caption{CappedRecall@25}
    \label{subfig:components_25}
  \end{subfigure}

  \caption{Number of components vs recall@k, combined across datasets. Comparison of three embedding models for the proposed method versus the baseline. Our method consistently performs better than the baseline for queries with more components. The proposed method consistently outperforms the baseline as the number of query components increases. At $k{=}25$ (Figure~\ref{subfig:components_25}), our method maintains stable performance for highly compositional queries, whereas the baseline exhibits a performance drop.}
  \label{fig:components-combined}
\end{figure}

\begin{figure}[th]
  \centering

  \begin{subfigure}{\linewidth}
    \centering
    \includesvg[width=0.7\linewidth]{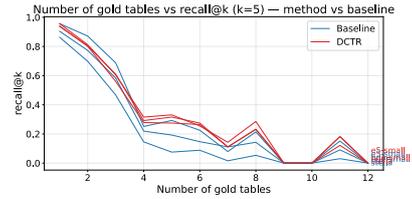}
    \caption{CappedRecall@5}
  \end{subfigure}

  \vspace{0.5em}

  \begin{subfigure}{\linewidth}
    \centering
    \includesvg[width=0.7\linewidth]{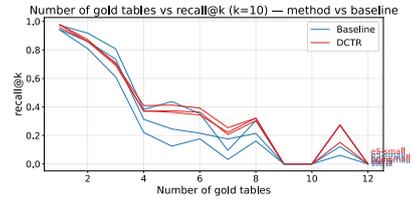}
    \caption{CappedRecall@10}
  \end{subfigure}

  \vspace{0.5em}

  \begin{subfigure}{\linewidth}
    \centering
    \includesvg[width=0.7\linewidth]{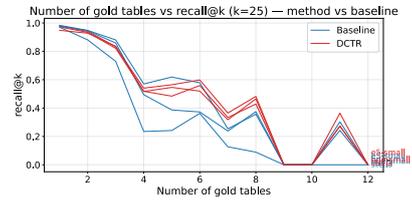}
    \caption{CappedRecall@25}
  \end{subfigure}

  \caption{Number of tables gold tables vs recall@k, combined across datasets. Comparison of three embedding models for the proposed method versus the baseline. Both methods exhibit a degradation in capped recall as there the number of gold tables increases. The proposed method behaves similarly to the baseline, indicating limited advantage when there is a lot of relevant tables to retrieve.}
  \label{fig:groupsize-combined}
\end{figure}

\begin{figure}[th]
  \centering

  \begin{subfigure}{\linewidth}
    \centering
    \includesvg[width=0.7\linewidth]{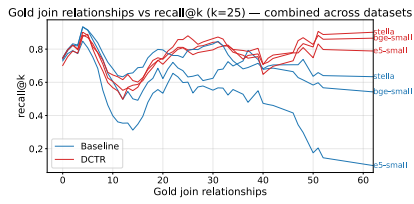}
    \caption{CappedRecall@5}
  \end{subfigure}

  \vspace{0.5em}

  \begin{subfigure}{\linewidth}
    \centering
    \includesvg[width=0.7\linewidth]{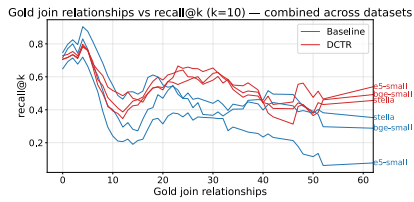}
    \caption{CappedRecall@10}
  \end{subfigure}

  \vspace{0.5em}

  \begin{subfigure}{\linewidth}
    \centering
    \includesvg[width=0.7\linewidth]{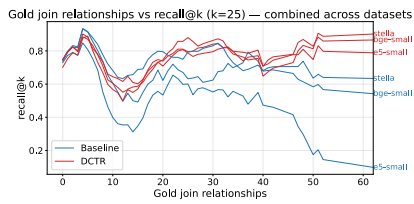}
    \caption{CappedRecall@25}
  \end{subfigure}

  \caption{Number of gold tables join relationships vs recall@k, combined across datasets. Comparison of three embedding models for the proposed method versus the baseline. DCTR improves performance on queries for which gold tables are densely connected.}
  \label{fig:joins-combined}
\end{figure}

Figures [\ref{fig:qlen-combined}, \ref{fig:components-combined}, \ref{fig:groupsize-combined}, \ref{fig:joins-combined}] report per-query analyses for both query and data complexity at $k=5$ and $k=10$.

In addition to the results discussed in the main text, Table~\ref{fig:groupsize-combined} presents recall as a function of the number of gold tables. As expected, retrieval difficulty increases with the number of required tables, leading to consistent recall degradation. Our method follows a similar trend to the dense baseline, indicating that large multi-table requirements remain challenging.

The remaining analyses replicate the trends observed in the main section. On the query complexity side, performance improves with increasing query length and number of extracted components. On the data complexity side, the method is more robust for queries involving densely connected tables, yet performance deteriorates when a large number of gold tables must be jointly retrieved.

\section{Datasets and Query distribution}
\label{appendix:distribution}

\begin{figure}[th]
  \centering

  \begin{subfigure}{\linewidth}
    \centering
    \includesvg[width=0.6\linewidth]{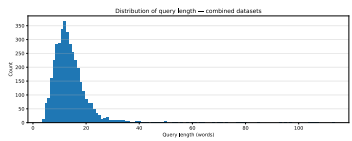}
    \caption{Distribution of query length}
  \end{subfigure}

  \vspace{0.5em}

  \begin{subfigure}{\linewidth}
    \centering
    \includesvg[width=0.6\linewidth]{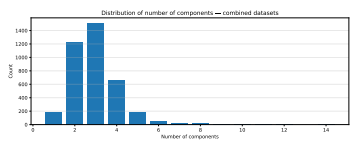}
    \caption{Distribution of number of extracted schema/value components}
  \end{subfigure}

  \vspace{0.5em}

  \begin{subfigure}{\linewidth}
    \centering
    \includesvg[width=0.6\linewidth]{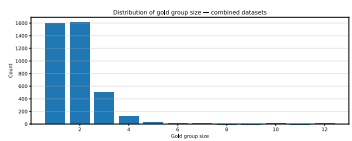}
    \caption{Distribution of average group size of gold tables}
  \end{subfigure}

  \begin{subfigure}{\linewidth}
    \centering
    \includesvg[width=0.6\linewidth]{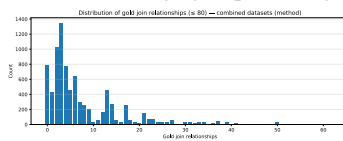}
    \caption{Distribution of number of join relationships in gold tables}
  \end{subfigure}

  \caption{Distribution of analyzed properties. Results combined for all datasets.}
  \label{fig:dist-combined}
\end{figure}

Figure~\ref{fig:dist-combined} shows the distribution of query and data complexity attributes across the evaluated queries. Most queries are shorter than 40 tokens, contain 1-6 extracted components, require 1-4 gold tables, and involve 1-10 join relationships among those tables. These distributions contextualize the per-query analyses in the main paper and clarify the typical structural characteristics of the evaluated workloads.

Additionally, Table~\ref{tab:dataset-characteristics} summarizes benchmark characteristics. BEAVER and FIBEN contain substantially more tables and higher average tables per database than BIRD, making retrieval considerably more challenging in these enterprise-scale settings. In contrast, BIRD exhibits the highest number of foreign keys per database, a structural property that better aligns with our connectivity-aware design and increases the likelihood of benefiting from group expansion.

\begin{table}[htbp]
  \centering
  \small
  \begin{tabular}{lrrr}
    \toprule
    \textbf{Metric}        & \textbf{BEAVER}     & \textbf{BIRD}       & \textbf{FIBEN}      \\
    \midrule
    Queries       & 209        & 1534       & 300        \\
    Unique DBs   & 6          & 11         & 1          \\
    Avg query\ (w)   & 31.9    & 14.6      & 12.3       \\
    Avg SQL (tok)& 152.3      & 25.4       & 61.9       \\
    Tables       & 463        & 75         & 152        \\
    Columns      & 4238       & 798        & 373        \\
    Avg Tab./DB    & 77.2       & 6.8        & 152.0      \\
    Avg Col./Tab.    & 9.2        & 10.6       & 2.5        \\
    Avg FK/Tab.     & 1.1        & 1.4        & 1.0        \\
    \bottomrule
  \end{tabular}
  \caption{Benchmark datasets characteristics.}
  \label{tab:dataset-characteristics}
\end{table}

\end{document}